%% file: Main.tex
\documentclass[5p,times]{elsarticle}

\usepackage{amsmath}
\usepackage{amssymb}
\usepackage{amsfonts}
\usepackage{multirow}
\usepackage{longtable}
\usepackage{caption}
\usepackage{subcaption}
\usepackage{pifont}
\usepackage{array}
\usepackage{url}
\newcolumntype{P}[1]{>{\centering\arraybackslash}p{#1}}

\begin{document}

\let\WriteBookmarks\relax
\def\floatpagepagefraction{1}
\def\textpagefraction{.001}



\title{Raijū: Reinforcement Learning-Guided Post-Exploitation for Automating Security Assessment of Network Systems}                      



%
\author[1,2]{Van-Hau Pham\corref{cor1}}%
\ead{haupv@uit.edu.vn}
\cortext[cor1]{Corresponding author}






\affiliation[1]{organization={University of Information Technology},
    city={Ho Chi Minh City},
    country={Vietnam}}
\affiliation[2]{organization={Vietnam National University},
    city={Ho Chi Minh City},
    country={Vietnam}}

\author[1,2]{Hien Do Hoang}
\ead{hiendh@uit.edu.vn}

\author[1,2]{Phan Thanh Trung}
\ead{19520321@uit.edu.vn}
\author[1,2]{Van Dinh Quoc}
\ead{19520240@uit.edu.vn}

\author[1,2]{Trong-Nghia To}
\ead{nghiatt@uit.edu.vn}

\author[1,2]{Phan The Duy}
\ead{duypt@uit.edu.vn}

\begin{abstract}
In order to assess the risks of a network system, it is important to investigate the behaviors of attackers after successful exploitation, which is called post-exploitation. Although there are various efficient tools supporting post-exploitation implementation, no application can automate this process. Most of the steps of this process are completed by experts who have profound knowledge of security, known as penetration testers or pen-testers. To this end, our study proposes the Raijū framework, a Reinforcement Learning (RL)-driven automation approach that assists pen-testers in quickly implementing the process of post-exploitation for security-level evaluation in network systems. We implement two RL algorithms, Advantage Actor-Critic (A2C) and Proximal Policy Optimization (PPO), to train specialized agents capable of making intelligent actions, which are Metasploit modules to automatically launch attacks of privileges escalation, gathering hashdump, and lateral movement. By leveraging RL, we aim to empower these agents with the ability to autonomously select and execute actions that can exploit vulnerabilities in target systems. This approach allows us to automate certain aspects of the penetration testing workflow, making it more efficient and responsive to emerging threats and vulnerabilities. The experiments are performed in four real environments with agents trained in thousands of episodes. The agents automatically select actions and launch attacks on the environments and achieve over 84\% of successful attacks with under 55 attack steps given. Moreover, the A2C algorithm has proved extremely effective in the selection of proper actions for automation of post-exploitation.
\end{abstract}

\begin{keyword}
Post Exploitation \sep Reinforcement Learning\sep Penetration Testing\sep Automation\sep Red Teaming
\end{keyword}




\maketitle

\input{1-Introduction}
\input{3-ReinforcementLearning}

\input{4-Design}

\input{5-Experiments}

\input{2-Relatedwork}

\input{6-Conclusion}









\bibliographystyle{unsrt}

\bibliography{refs.bib}

\end{document}

%% file: 1-Introduction.tex
\section{Introduction} \label{sec_introduction}


 With the growing reliance on technology, the threat of cyber-attacks has surged, demanding robust security measures. In this context, penetration testing (PT) and red team assessments are crucial components of modern cybersecurity \cite{PT_redteam_overview}. The term of PT, often referred to as ethical hacking, is a systematic approach that involves skilled testers using tools and manual techniques to uncover vulnerabilities in networks, applications, systems, and more. This method provides organizations with a comprehensive matrix of vulnerabilities and patching issues, making it an effective means of assessing existing security weaknesses. However, it is worth noting that real attackers do not simply rely on vulnerability scans as penetration testers do; they focus on post-exploitation behaviors after successfully compromising a target. In contrast, red team testing goes beyond the identification of vulnerabilities. It simulates the tactics and behaviors that attackers employ after gaining access to a system. Red teaming requires testers with diverse skills to pursue specific objectives, such as accessing particular data or systems. This approach helps organizations assess the potential negative impact and losses that could result from a breach. In the endless journey of cybersecurity, both penetration testing and red team assessments play pivotal roles in safeguarding against evolving threats and ensuring the resilience of an organization's defenses \cite{teichmann2023evolution}.

In PT, post-exploitation refers to the stage of a cyberattack occurring after intruders successfully gain unauthorized access to a network system or application.  The post-exploitation actions include lateral movement, privilege escalation, information gathering, and the creation of backdoors, among others. To simulate the actions of an actual attacker, these procedures are carried out as covertly as feasible. Additionally, this phase enables testers to record methods used to access assets and propose ways to protect the systems or networks against unauthorized access. Many previous studies have used this aspect of post-exploitation to construct automatic PT techniques to make the PT procedure more comfortable \cite{Kujanp_automating_privilege_escalation_rl,Ryusei_automating_post_exploitation_rl}.

To achieve success in the post-exploitation stage, experts with practice skills and in-depth security knowledge are required to execute exploits. Although there are tools supporting this stage, such as Empire \cite{empire}, DeepExploit \cite{deep_exploit}, Metasploit \cite{metasploit}, and OpenVAS \cite{openvas}, testers have to take the full force and time consumption for this process. However, the automation of these tools depends on experts customizing the configurations and environments of the tools and deciding some steps in the testing process. Many advantages can be achieved for defenders if this process is fully automatic, especially in cases of increasingly sophisticated attacks. As a result, many attacks can be predicted and detected so that the defenders can make appropriate responses immediately with automatic and accurate support.

To simulate human intelligence in the process of automation security assessment, the application of RL is sufficient consideration \cite{ghanem2023hierarchical}. A RL-based agent is trained by enormous amounts of attack data to perform suitable actions in a target environment. Through a PT tool, intelligent actions can be carried out to achieve certain objectives and evade detection in the post-exploitation stage.

Recently, there have been studies focusing on applying machine learning (ML) or deep learning (DL) in the detection of malware, phishing, or intrusion \cite{Apruzzese_ml_dl_cyber_security} \cite{Cavusoglu_ml_ids} \cite{Cui_dl_malware}. These studies utilized prepared datasets for the learning process using two popular types of ML that are supervised learning and unsupervised learning. However, preparing a dataset for an environment that is continuous and real-time, such as post-exploitation, becomes a considerable problem. Therefore, supervised and unsupervised learning is not a suitable choice for post-exploitation automation. RL is a proper type of ML for the automation of post-exploitation. The RL agent can be trained from states that update from complex and real-time environments so that it can give actions continuously. 

Kalle et al. \cite{Kujanp_automating_privilege_escalation_rl} proposed a method leveraging RL to perform privilege escalation attacks automatically in the Windows 7 environment. Another research \cite{Ryusei_automating_post_exploitation_rl} combined RL and the PowerShell Empire tool to build a method for the automation of post-exploitation. The RL agent selects one of the modules of the tool as an action and then executes the exploit in the environment. The experience shows that the agent can give optimal actions and the privilege of the administrator can be obtained. Moreover, in the study of Yi et al. \cite{Yi_DRL_pentesting}, the MulVAL attack graph is integrated with the DDQN algorithm to enhance rewards using the prior knowledge from the attack graph. The experiments demonstrated that the MulVAL DDQN algorithm (MDDQN) accelerated the convergence speed and substantially improved the efficiency of attack path planning. The outcomes highlighted the increasingly evident benefits of the MDDQN algorithm as the experimental scenarios became more complex.

From a contextual perspective, Mohamed et al. \cite{ghanem2023hierarchical} presented the Intelligent Automated Penetration Testing Framework (IAPTF), revolutionizing PT by treating it as a partially observed Markov decision process (POMDP). This innovative framework leverages Hierarchical deep RL to automate the decision making that effectively tackles the formidable challenge of solving large POMDPs within extensive network environments. Remarkably, the scalability of IAPTF is directly proportional to the network's size, making it highly adaptable. Beyond this, IAPTF streamlines the process of retesting similar networks, a frequent occurrence in real-world PT scenarios, greatly enhancing its practical utility and efficiency. 

In another study, the authors introduced an innovative architecture, named deep Cascaded Reinforcement Learning Agents (CRLA) \cite{tran2022cascaded}, designed to tackle the challenges posed by large discrete action spaces in autonomous PT simulations. With an algebraic action decomposition, CRLA succeeds at identifying optimal attack policies in complex cybersecurity networks, outperforming traditional deep Q-learning agents commonly employed in autonomous PT. The study validates CRLA's superior performance on simulated CybORG scenarios, showcasing its competitive advantage over single DDQN agents. Furthermore, CRLA's scalability to larger action spaces with sub-linear computational complexity highlights its potential for real-world application.

While previous research has made valuable contributions to the advancement of automated PT methods, these related studies predominantly focus on isolated individual aspects, neglecting potential interconnections that could enhance flexibility. In this study, we focus on leveraging the potential of ML to strengthen network security. Specifically, we dig into the realm of RL-driven automation for the post-exploitation process within networks and systems. By taking advantage of RL algorithms, namely Advantage Actor-Critic (A2C) and Proximal Policy Optimization (PPO), we aim to empower intelligent agents to navigate complex security landscapes. These agents, equipped with insights from the algorithms, are primed to assess server states and identify potential vulnerabilities. The main contributions of this paper are as follows.
\begin{itemize}
    \item Propose Raijū framework which has an intelligent RL agent to select proper Metasploit modules as actions for automatic exploit execution in the post-exploitation phase.
    \item Employ and compare two types of RL algorithms, including A2C and PPO in Raijū for seeking the paths and post-exploitation actions for widely spreading the control in the network system.
    \item Perform various experiments to evaluate the effectiveness of our framework on real-world vulnerable Windows and Linux machines to perform attacks related to privilege escalation, gathering hashdump, and lateral movement.
\end{itemize}

The remainder of this work is organized as follows. The background of RL is provided in \textbf{Section~\ref{sec_rl_background}}. \textbf{Section~\ref{sec_methodology}} describes the architecture of our Raijū framework in the context of conducting PT of the network systems. The experimental results are given in \textbf{Section~\ref{sec_experiment}}. We briefly give an overview of related works on RL-based exploitation in \textbf{Section~\ref{sect_relatedworks}}. Finally, in \textbf{Section~\ref{sec_conclusion}}, we conclude our research.

%% file: 3-ReinforcementLearning.tex
\section{Background} \label{sec_rl_background}
This section briefly introduces an overview of RL and their types of learning and decision-making in dynamic environments.
\subsection{Reinforcement Learning}

RL \cite{RL_2,RL} is a dynamic field within ML that focuses on training agents to make sequential decisions by interacting with an environment. In the context of RL, there are several distinct types that characterize different approaches to learning and decision-making. Understanding these types is essential for comprehending the diversity of techniques used in various applications.

Types of RL are categorized as follows:
\begin{itemize}
    \item \textbf{Model-Based RL:} In this approach, agents construct an internal model of the environment to predict future states and rewards. This model guides decision-making, allowing agents to plan ahead. Model-based RL is often beneficial in scenarios where the environment is complex or costly to interact with directly.

    \item \textbf{Model-Free RL:} Unlike model-based approaches, model-free RL directly learns optimal actions without constructing an explicit model. It relies on trial-and-error learning, updating its knowledge based on the observed rewards. This type is especially useful when the environment is highly dynamic or lacks a well-defined model.

    \item \textbf{Value-Based RL:} Value-based RL aims to learn the value function, which estimates the expected cumulative reward from a given state. Algorithms like Q-learning and Deep Q-Networks (DQN) \cite{DQN} fall under this category. Value-based methods are well-suited for tasks with discrete action spaces.

    \item \textbf{Policy-Based RL:} Policy-based RL revolves around directly learning the optimal policy that dictates an agent's actions. It is particularly effective for continuous action spaces and can handle stochastic policies more naturally.

    \item \textbf{Actor-Critic RL:} This type combines elements of both value-based and policy-based approaches. An actor selects actions based on a learned policy, while a critic evaluates the value of those actions. Actor-critic methods strike a balance between exploration and exploitation.

    \item \textbf{Proximal Policy Optimization:} PPO is a popular policy optimization algorithm that focuses on improving the stability and efficiency of policy learning. It addresses some of the challenges posed by traditional policy gradient methods.
\end{itemize}

\subsection{Advantage Actor Critic - A2C}

A2C \cite{A2C,A2C_PPO} is a popular RL algorithm that combines elements of both policy-based and value-based approaches. A2C is designed to address some of the limitations of traditional policy gradient methods, offering improved stability and sample efficiency. More specifics, it employs an actor-critic architecture, where the actor generates actions based on a learned policy, and the critic estimates the value of these actions. This separation allows A2C to simultaneously learn both optimal actions and value estimates. Therein, a key innovation of A2C is the introduction of the advantage function. The advantage function measures the relative value of taking a particular action compared to the average action value. It provides a more informative signal for policy updates, resulting in faster convergence and reduced variance. A2C supports asynchronous updates, where multiple instances of the environment run in parallel, each with its own actor and critic. This parallelism enhances sample efficiency and accelerates learning, making A2C particularly suitable for environments with high-dimensional state and action spaces. In addition, A2C employs policy gradient optimization to update the actor's policy. By utilizing the advantage function to scale the gradient, A2C mitigates issues like high variance commonly associated with policy gradients. 
The gradient of A2C can be  expressed as \textbf{(\ref{equation:A2C})}:

\begin{equation}
\label{equation:A2C}
\fontsize{9}{9}\selectfont
\nabla_\theta J(\theta) = \mathbb{E}_{r\sim p}(\tau|\theta) \left[ \sum_{t=0}^{T-1}\nabla _{\theta} 
  \log \pi_{\theta}(a_t|s_t)A^{\pi_{\theta}}(s_t,a_t)\right] 
\end{equation}

where  $A^{\pi_{\theta}}(s_t,a_t)$ is the estimated advantage following policy $\pi_{\theta}$.

In conclusion, A2C offers a robust and versatile approach to RL by combining policy-based and value-based techniques. Its advantages include improved stability, faster convergence, and suitability for parallel environments. A2C finds applications across a diverse range of domains, making it a valuable tool for training agents to make optimal decisions in complex and dynamic environments.

\begin{figure*}[!t]
    \centering
    \includegraphics[width=0.8\textwidth]{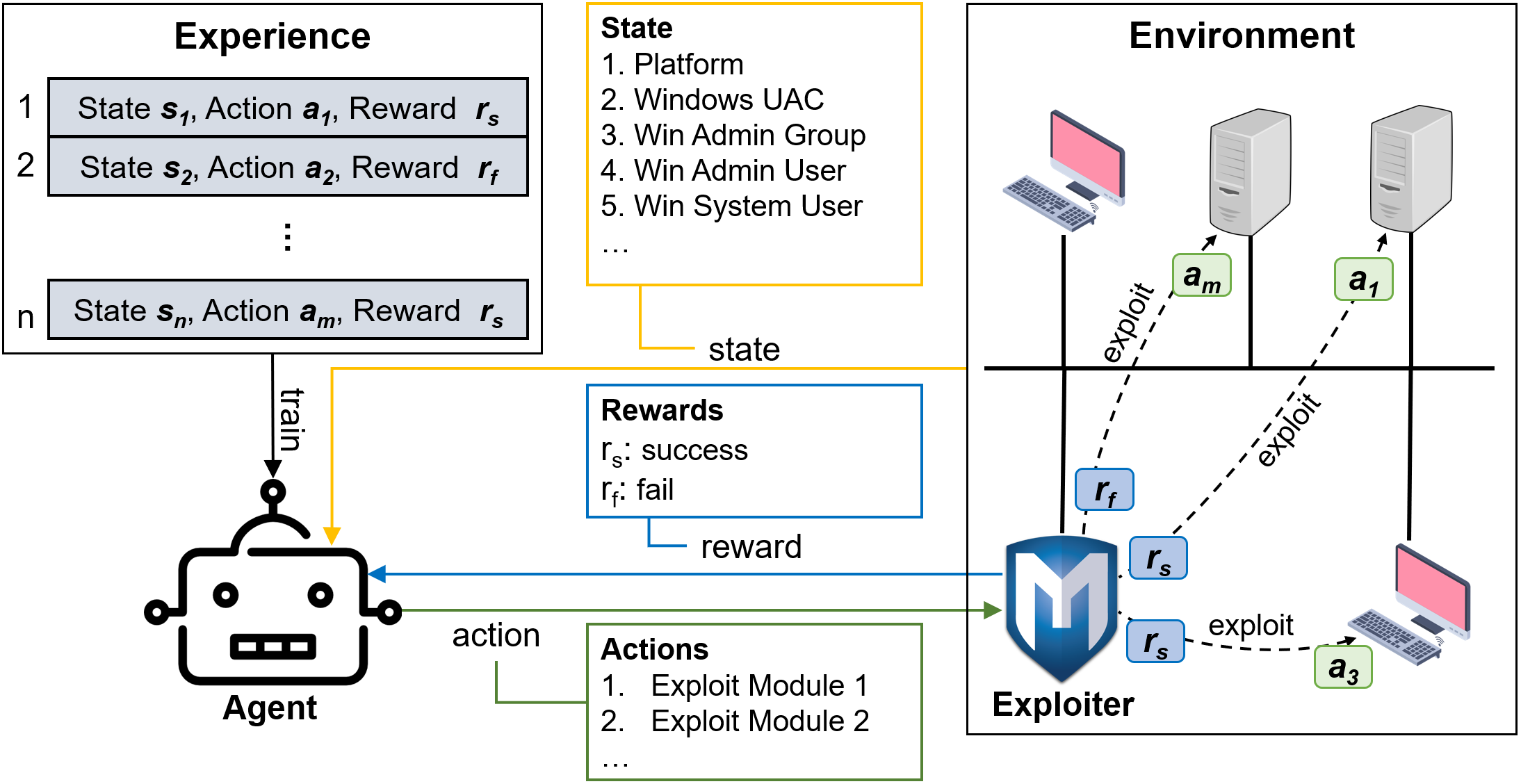}
    \caption{The overview of Raijū architecture}
    \label{fig_reinforcement_learning_Raiju}
\end{figure*}

\subsection{Proximal Policy Optimization - PPO}

PPO \cite{PPO,A2C_PPO} is an advanced RL algorithm that offers several advantages over traditional policy optimization methods. These advantages contribute to its popularity and effectiveness in training agents for complex tasks. Firstly, PPO is designed with stability in mind. It includes a clipping mechanism that limits the magnitude of policy updates during optimization. This prevents drastic policy changes that could lead to divergence or unstable learning. The controlled updates enhance the algorithm's robustness and make it less sensitive to hyperparameter choices. Secondly, PPO efficiently uses collected experience to update policies. By reusing sampled data multiple times within a single iteration, PPO reduces the variance of updates, making better use of the collected experience. This sample efficiency is crucial for practical applications, especially in scenarios where collecting data is resource-intensive or time-consuming. Thirdly, PPO is versatile and can be employed with different neural network architectures, such as feedforward networks or recurrent networks. This compatibility allows practitioners to tailor the algorithm to their specific problem domains and leverage network structures that suit their tasks. Moreover, PPO can handle both continuous and discrete action spaces, making it versatile for a wide range of tasks. This adaptability is particularly beneficial in domains where the nature of actions varies. We examine the fundamental version of PPO-Clip, utilizing the surrogate objective \textbf{(\ref{equation:PPO}, \ref{equation:PPO_2})} for each timestep $t$: 
\begin{equation}
\label{equation:PPO}
    \mathcal{L}_{CLIP} = \mathbb{E}_{r\sim p}(\tau|\theta_{k}) \times  Z
\end{equation}

\begin{equation}
\label{equation:PPO_2}
Z = \left[ \sum_{t=0}^{T-1}\left[ \min (r_t(\theta), clip(r_t(\theta), 1 - \epsilon, 1 + \epsilon )) \mathrm{A}_{t}^{\pi_{\theta_{k}}} \right] \right]
\end{equation}

where $r_t(\theta) = \frac{\pi_\theta (a_t|s_t)}{\pi_{\theta_k} (a_t|s_t)}$ and $\epsilon$ is a narrow threshold that effectively constrains the difference between the new policy and the old policy.

In summary, PPO offers a combination of stability, robustness, sample efficiency, and controlled exploration. These advantages contribute to its effectiveness in training RL agents across various domains, and make it a popular choice for practitioners seeking reliable and efficient algorithms for complex tasks.

\subsection{Comparing A2C and PPO}
A2C uses an objective function called the cumulative objective function, while PPO employs an objective function referred to as the surrogate objective function. The cumulative objective function in A2C provides a way to guide learning by optimizing the total reward over a certain time horizon, whereas the surrogate objective function in PPO provides a way to guide learning by optimizing the reward at each step. 

Furthermore, the differences between the two algorithms are also reflected in:
\begin{itemize}
    \item \textbf{Update mechanism:} In A2C, updates are performed after each time step, while PPO typically performs multiple update steps on a large batch of collected data.
    \item \textbf{Primary objective:} A2C focuses on optimizing the advantage, which is the difference between the predicted policy value and the state value. PPO focuses on optimizing the surrogate objective, using a loss function defined by combining the primary objective and penalty terms.
    \item \textbf{Constraint:} PPO uses a KL-divergence constraint to limit policy changes, ensuring that updates are not too large. In contrast, A2C does not apply this constraint and often requires other control measures to ensure stability in the learning process.
    \item \textbf{Performance:} PPO typically achieves more stable performance compared to A2C. The use of the KL-divergence constraint in PPO reduces the variance of policy updates and helps optimize progress in a more stable manner.
    \item \textbf{Parallelizability:} Because PPO performs multiple update steps on a large batch of data, it is suitable for running on multiple threads or concurrently in multiple environments to accelerate the learning process. On the other hand, A2C can be parallelized but may not be as effective as PPO.
\end{itemize}

%% file: 4-Design.tex
\section{Methodology} \label{sec_methodology}
\subsection{Problem Statement}
To address the above-mentioned problem, we leverage RL to propose Raijū framework for enhancing the automation of post-exploitation. The name Raijū is inspired by the meaning of "thunder beast", which is a legendary creature from Japanese mythology. In our framework, an RL agent is trained by data gathered from real-time environments, and then this agent collects necessary information as a state from a network environment and gives a suitable action. Based on the given action, support tools can execute exploits to the target for achieving the goals. In this paper, our agent acts as a red teaming tester with the aim of privilege escalation, hashdump (hash of credential) collection, and lateral movement in networks. To reach these goals, it is necessary to determine the following aspects.
\begin{itemize}
    \item The variability of target environments for feature selection to build agent states.
    \item Actions to interact with the environment for the post-exploitation process.
    \item Mechanism of reward measurement for optimizing results of exploit executions from the action given.
\end{itemize}

\subsection{Overview}

This section describes our proposed Raijū framework for automating the post-exploitation stage. The architecture of the Raijū framework is depicted in \textbf{Fig.~\ref{fig_reinforcement_learning_Raiju}}.  We choose variables changing when an event occurs in an environment as features of the state \textit{s}. Then, exploit modules are selected for the determination of RL actions \textit{a}. Finally, the reward \textit{r} is defined to obtain maximum effectiveness from the action. 

In the more detailed level, the agent is provided a set of $s$, $a$, and $r$ as the experience collected from environments for the learning process. Various environments are built for experienced collections to improve the accuracy of the agent. In addition, an exploiter is built into the design to launch exploits in environments. The exploiter receives an action $a$ from the agent, then responds with a corresponding reward $r$ back to the agent based on the success or failure of the exploit.

Furthermore, we explore the creation of computational models that make use of the strengths of the well-known RL algorithms, including A2C and PPO. For A2C, the loss function can be expressed as \textbf{(\ref{equation:Loss function A2C})}:

\begin{equation}
\label{equation:Loss function A2C}
\fontsize{9}{9}\selectfont
\mathcal{L}_{A2C} = \mathcal{L}_{actor} + \mathcal{L}_{critic} - \beta \times entropy
\end{equation}
This formulation entails the following components:
\begin{itemize}
    \item  $\mathcal{L}_{actor}$ represents the actor loss, aiming to maximize the expected cumulative reward by updating the policy (actor-network).
    \item  $\mathcal{L}_{critic}$ represents the critic loss, aiming to minimize the mean squared error (MSE) between predicted state values and actual rewards, updating the value function (critic network).
    \item  \(\beta\) is a hyperparameter, often a small value, controlling the influence of the entropy regularization term.
    \item  $entropy$ represents the entropy of the action distribution, encouraging exploration.
\end{itemize}

The actor loss, critic loss, and entropy value components are calculated as follows the \textbf{Formula \ref{equation:Actor_A2C}, \ref{equation:Critic_A2C}, \ref{equation:entropy_A2C}}:
\begin{equation}
\label{equation:Actor_A2C}
    \mathcal{L}_{actor} = -\log(\pi_\theta(a|s)) \times advantage
\end{equation}

\begin{equation}
\label{equation:Critic_A2C}
    \mathcal{L}_{critic} = MSE(V(s), R_t)
\end{equation}

\begin{equation}
\label{equation:entropy_A2C}
\fontsize{9}{9}\selectfont
 \text{entropy} = -\sum \pi_\theta(a|s) \log(\pi_\theta(a|s))
\end{equation}

where:
\begin{itemize}
   \item $\pi_\theta(a|s)$ is the probability of taking action $a$ given state $s$ according to the policy (actor network).
   \item $advantage$ is the advantage, which is a measure of how much better the action taken was compared to the expected value in that state.
   \item  $V(s)$ is the predicted state value by the critic network.
   \item  $R_t$ is the cumulative reward or return.
\end{itemize}

In terms of PPO, the objective function $\mathcal{L}_{PP0}$ is structured around two crucial elements: the clipped surrogate objective, computed as shown in \textbf{(\ref{equation:PPO})} and \textbf{(\ref{equation:PPO_2})}, and the value function loss, denoted as  $\mathcal{L}_{VF}$. The value function loss is the MSE between the predicted value of the state and the discounted sum of rewards. More specifically, the calculation formula is represented by  \textbf{(\ref{equation:Value_loss_PPO})}: 
\begin{equation}
\label{equation:Value_loss_PPO}
   \mathcal{L}_{VF} = \frac{1}{2}\mathcal{E}\left[\left(V_\theta(s_t) - (R_t - V_\theta(s_t))^2\right)\right]
\end{equation}

here, $V_\theta(s_t)$ is the predicted value of the state. In essence, the overall objective function in PPO combines these components:
\begin{equation}
\label{equation:PPO_overalll}
\mathcal{L}_{PP0} = \mathcal{L}_{CLIP}(\theta) - c_1 \times \mathcal{L}^{VF}(\theta) + c_2 \times S[\pi_\theta](s_t)
\end{equation}
Within this formula, \( c_1 \) and \( c_2 \) are hyperparameters, and \( S[\pi_\theta](s_t) \) is the entropy of the policy, encouraging exploration.

\subsection{Environment Definition}

Environments for learning and testing processes include machines that run on Windows and Linux platforms and contain built-in and popularly installed applications and services. Popular ports are also open for the post-exploitation phase. For the Windows platform, we select Windows 7 because various vulnerabilities are available on this platform. Besides, Metasploitable 2 is chosen as the Linux platform since it contains popular vulnerabilities by default, so it is not necessary to customize it.

In our methodology, we utilize Metasploit and Nmap as adjunctive instruments for information acquisition in both training and testing phases. To ensure the effectiveness of an exploit, we meticulously oversee the commencement of new sessions within Metasploit and actively engage with these sessions following the execution of an exploit. This procedure enhances the precision of our exploit outcome assessment and augments the overall efficacy of our approach.

In the training stage, directly interacting with the environment takes much time and resource consumption and causes unexpected connection errors, which occur when training episodes reach thousands. To tackle these problems, we save the values of environments and the results of exploits into CSV files. This idea is based on the works \cite{Kujanp_automating_privilege_escalation_rl} \cite{Hu_auto_pentest}, which put saved input for training instead of directly interacting with real environments. This approach is suitable because the agent can learn over states which are values achieved in the environments. Input for the agent includes the following components:
\begin{itemize}
    \item \textbf{Action Space:} The list of exploit modules that are selected from Metasploit modules.
    \item \textbf{State:} Values returned from an environment after successfully exploiting and controlling the compromised machine. These values are extracted from the Metasploit and Nmap.
    \item \textbf{Result:} The result of executing the selected exploit module.
    \item \textbf{Peer information:} Detail information of neighbor machines found from the compromised machine.
    \item \textbf{Target:} The condition the agent must achieve when performing an attack. It is also considered a stop condition for execution. This target can differ from attack to attack, ranging from escalated privileges, and obtained hashdump to successful lateral movement.
\end{itemize}

\subsection{Agent State}

The state of the agent is defined by 10 features, which are shown in \textbf{Table~\ref{tab:rl_agent_state}}. Values of these features are obtained from real environments. When an attack is launched on a host, these values are obtained by the environment that sends them to the agent. \textit{Platform} is an important feature for the proper action decision that gives a correct exploit module for Windows or Linux platforms. \textit{Windows UAC}, \textit{Windows Admin Group}, \textit{Windows Admin User}, and \textit{Windows System User} are features to identify the permissions of the execution user on a compromised Windows machine. In the same way, the \textit{Linux Root User} reveals whether a compromised user on a Linux machine is the root user or not. Besides, since Linux kernels are open sources, there are vulnerabilities in some special versions of kernels and assign value for \textit{Linux Kernel Vul}. \textit{Hashdump} is the key factor for gaining access to a machine and includes username and password hashed for confidentiality insurance. \textit{Number of peers}, \textit{Peers index}, and \textit{Peer platform} consist of essential information for lateral movement attacks from the compromised machine.

\begin{table}[!t]
    \centering
    \caption{Feature definition of the agent state}
    \def\arraystretch{1.3}%
    \begin{tabular}{p{2cm}p{6cm}}
    \hline
    \textbf{Feature name} & \textbf{Description} \\
    \hline
    Platform & Operating system of a machine (0: Windows, 1: Linux, or -1: Unknown) \\
    \hline
    Windows UAC & The User Account Control (UAC) which is a Windows security feature (0: Enabled, 1: Disabled, -1: Unknown) \\
    \hline
    Windows Admin Group & Whether a compromised Windows user is in Administrator groups or not (0: Yes, 1: No, -1: Unknown) \\
    \hline
    Windows Admin User & Whether a compromised Windows user is an Administrator or not (0: Yes, 1: No, -1: Unknown) \\
    \hline
    Windows System User & Whether a compromised Windows user is NT SYSTEM or not (0: Yes, 1: No, -1: Unknown) \\
    \hline
    Linux Root User & Whether a compromised Linux user is the root user or not (0: Yes, 1: No, -1: Unknown) \\
    \hline
    Linux Kernel Vul & Whether a Linux kernel has vulnerabilities or not (0: No, 1: Yes, -1: Unknown) \\
    \hline
    Hashdump & Whether the hash of credential is gathered or not (0: No, 1: Yes) \\
    \hline
    Number of peers & The number of neighbor machines found from compromised machine \\
    \hline
    Peers index & The ID of a machine in the list of the found machines \\
    \hline
    Peer platform & Operating system of a peer (0: Windows, 1: Linux, or -1: Unknown) \\
    \hline
    \end{tabular}
\label{tab:rl_agent_state}
\end{table}

\subsection{Action Space}
We define RL actions that are real exploit modules used for the automation of the post-exploitation process in Metasploit. Although Metasploit supports many modules for exploits, we choose 99 well-known modules for the actions of the agent. These actions are categorized into 4 groups, shown in \textbf{Table~\ref{tab:classification_metasploit_module}}. There are 95 modules for privilege escalation attacks on both Windows and Linux because these platforms have various existing vulnerabilities. In addition, two modules are responsible for gathering hashdump and two modules launch lateral movement attacks through the SMB service.

\begin{table}[!b]
    \centering
    \caption{Classification of Metasploit modules for post-exploitation}
    \def\arraystretch{1.3}%
    \begin{tabular}{P{1.4cm}p{6.5cm}}
    \hline
    \textbf{Number of modules} & \textbf{Description} \\
    \hline
    23 & Privilege escalation on Windows \\
    \hline
    72 & Privilege escalation on Linux \\
    \hline
    2 & Hashdump collection on Windows and Linux platforms \\
    \hline
    2 & Exploiting SMB service to spread to other machines \\
    \hline
    \end{tabular}
\label{tab:classification_metasploit_module}
\end{table}

\subsection{Reward}

This section structures the reward mechanism for the Raijū agent. The reward of the action is calculated depending on the execution result and the loss of action. There are two types of rewards that the Raijū agent can receive. If the given action can successfully exploit a vulnerability by Metasploit, the agent receives the reward $r_s$, otherwise, the agent receives the reward $r_f$.

We use the Discounted Sum of Rewards equation to calculate the cumulative reward $R_t$ by summing the rewards, each multiplied by a discount factor based on their time step.

\begin{equation}
\label{equation:Reward Calculate}
R_t = \sum_{i=t}^{T} \gamma^{i-t} \cdot r_i
\end{equation}\\

There are three principal components:
\begin{itemize}
    \item $r_i$ signifies the reward obtained at time step \(i\).
    \item $\gamma$ is the discount factor, which falls within the range $[0, 1]$.
    \item $t$ denotes the current time step, $T$ indicates the end of an episode, $i$ iterates from $t$ to $T$ inclusively, representing the time steps within the episode.
\end{itemize}

The primary objective of this equation is to estimate the total expected reward that the agent can accumulate starting from the current time step until the conclusion of the episode. The value of the discount factor $\gamma$ plays a pivotal role in shaping the contribution of future rewards within this accumulation process. A smaller $\gamma$ assigns a higher weight to immediate rewards, while a larger $\gamma$ places greater emphasis on rewards that are expected in the future.

Each reward $r_i$ is attenuated by a discount factor $\gamma^{i-t}$, capturing the concept of diminishing influence of rewards as we project further into the future. This concept aligns with the premise that immediate rewards hold greater significance for the agent in comparison to rewards obtained in the distant future.

%% file: 5-Experiments.tex
\section{Experiments and Evaluation} \label{sec_experiment}
\subsection{Experimental settings}
Experiments are performed on a machine running on Ubuntu 20.04 with 16GB of memory and Python v3.7, Metasploit Framework v6.3, Pytorch v2.0, and Pymetasploit3 installed.

We set up four environments including ENV1, ENV2, ENV3, and ENV4 for experiments. Each environment consists of a compromised machine and some neighbor machines. Machines in an environment contain vulnerabilities that allow launch attacks of privilege escalation (PE), gathering hashdump (GH), and lateral movement (LM). Two platforms that are Windows 7 for Windows and Metaploitable (Kernel version 4.15) for Linux are used. Port 445 is used for LM attacks, called LM port, and value -1 indicates that there is no knowledge about port 445, which is opened whether on the machine or not. Details of the four environments are shown in \textbf{Table \ref{tab:experimantal_env}}.

\begin{table}[t]
    \centering
    \caption{The experimental environments}
    \def\arraystretch{1.3}%
    \begin{tabular}{cP{1.8cm}P{1.6cm}c}
    \hline
    \textbf{Environment} & \textbf{Compromised machine} & \textbf{Neighbor machines} & \textbf{LM Port}\\
    \hline
    \multirow{4}{*}{ENV1} &  \multirow{4}{*}{Windows} & Linux & -1 \\
    \cline{3-4}
    & & Windows & 445 \\
    \cline{3-4}
    & & Linux & 445\\
    \cline{3-4}
    & & Windows & -1 \\
    \hline
    \multirow{2}{*}{ENV2} &  \multirow{2}{*}{Windows} & Linux & -1 \\
    \cline{3-4}
    & & Windows & 445 \\
    \hline
    \multirow{3}{*}{ENV3} &  \multirow{3}{*}{Linux} & Windows & -1 \\
    \cline{3-4}
    & & Linux & 445 \\
    \cline{3-4}
    & & Windows & 445\\
    \hline
    \multirow{3}{*}{ENV4} &  \multirow{3}{*}{Linux} & Windows & -1 \\
    \cline{3-4}
    & & Windows & -1 \\
    \cline{3-4}
    & & Linux & 445\\
    \hline
    \end{tabular}
\label{tab:experimantal_env}
\end{table}

\subsection{Evaluation metrics and scenarios} \label{metrics_scenarios}
\subsubsection{Training parameters}
The architecture for A2C and PPO algorithms is built based on the Actor-Critic approach with a 3-layer neural network illustrated in \textbf{Table~\ref{tab:actor_critic_architecture}}.


\begin{table}[!b]
    \centering
    \caption{The network architecture of Actor-Critic}
    \def\arraystretch{1.3}%
    \begin{tabular}{cp{3.2cm}p{3.2cm}}
    \hline
    \textbf{Layer} & \textbf{Actor} & \textbf{Critic} \\
    \hline
    \textbf{Input} & The input dimension equal to the number of features of the state & The input dimension equal to the number of features of the state \\
    \hline
    \textbf{Hidden} & 256 units & 256 units \\
    \hline
    \textbf{Output} & The units equal the number of actions & One unit, which is the estimated value of a state\\
    \hline
    \end{tabular}
\label{tab:actor_critic_architecture}
\end{table}

In addition, the hyperparameter setting for A2C and PPO algorithms is shown in \textbf{Table \ref{tab:hyperparameter_setting}}. These hyperparameters have been carefully selected to strike a balance between exploration and exploitation while training the RL agents. The discount factor of 0.99 encourages the agents to consider long-term rewards, while the learning rates for the Actor and Critic networks, set to 0.0003 and 0.001 respectively, determine the speed at which the agents adapt their strategies. For the PPO algorithm, an epsilon clip of 0.2 is applied to limit the policy updates, promoting more stable training.

\begin{table}[!b]
    \centering
    \caption{Hyperparameter settings}
    \def\arraystretch{1.3}%
    \begin{tabular}{p{3.5cm}P{1.5cm}P{1.5cm}}
    \hline
    \textbf{Hyperparameter} & \textbf{Value} & \textbf{Algorithm} \\
    \hline
    Discount factor & 0.99 & A2C, PPO \\
    \hline
    Learning rate for Actor & 0.0003 & A2C, PPO \\
    \hline
    Learning rate for Critic & 0.001 & A2C, PPO \\
    \hline
    Epsilon clip & 0.2 & PPO \\
    \hline
    \end{tabular}
\label{tab:hyperparameter_setting}
\end{table}

\subsubsection{Evaluation criteria} \label{sec_evaluation_criteria}

Results are measured in 100 tests in the four environments at various numbers of training episodes and the agent performed 4000 steps at maximum to reach the target. Three types of attacks including PE, GH, and LM are launched. For PE and GH attacks, we count the number of successful executions and the average figure of performed actions. In cases of LM attack, since most tests witness at least one neighbor machine exploited from the compromised machine, we only consider tests where all the neighbor machines in an environment are exploited from the compromised machine.

In experiments, we measure the following metrics to evaluate the effectiveness of reward measurement, the number of episodes, and algorithms.

\begin{itemize}
    \item SUCC-PE: The number of tests successfully launching attacks of PE of the total tests.
    \item SUCC-GH: The number of tests successfully launching attacks of GH of the total tests.
    \item SUCC-LM: The number of tests successfully launching attacks of LM of the total tests. It only counts tests in which attacks launched from the compromised machine successfully exploit all the neighbor machines in an environment.
    \item AVG-PE: The average of actions that the agent uses for a successful attack execution of PE.
    \item AVG-GH: The average of actions that the agent uses for a successful attack execution of GH.

    \item FAIL-LM: The number of tests that fail when launching LM attacks of the total tests.

\end{itemize}
We do not measure the average of actions that the agent uses for a successful attack execution of LM because this type of attack is extremely complex. It is necessary to try and repeat many steps to gain a successful attack, so averaging this criterion does not make sense.

\begin{figure}[!b]
    \centering
    \includegraphics[width=0.4\textwidth]{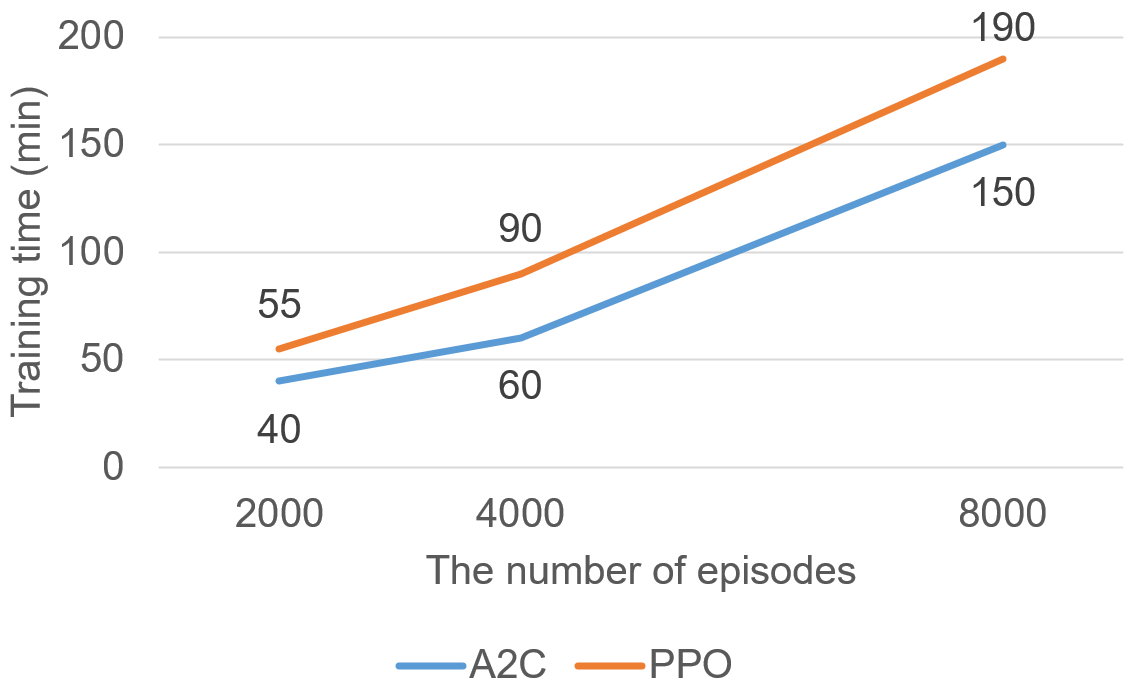}
    \caption{Time consumption for training RL models with various episode numbers.}
    \label{time_consumption_training}
\end{figure}

\begin{figure}[!t] 
\centering
\begin{tabular}{cc}
\subfloat[\textbf{A2C Algorithm}]
{
	\includegraphics[width=0.46\textwidth]{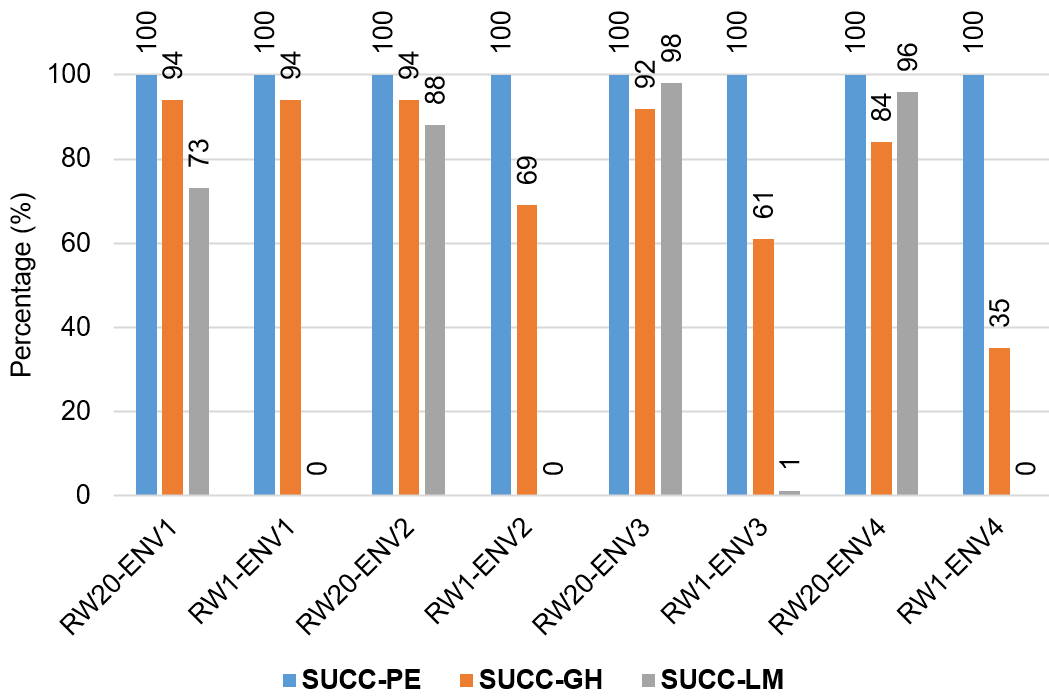}
	\label{}
}\\%
\subfloat[\textbf{PPO Algorithm}]
{
	\includegraphics[width=0.46\textwidth]{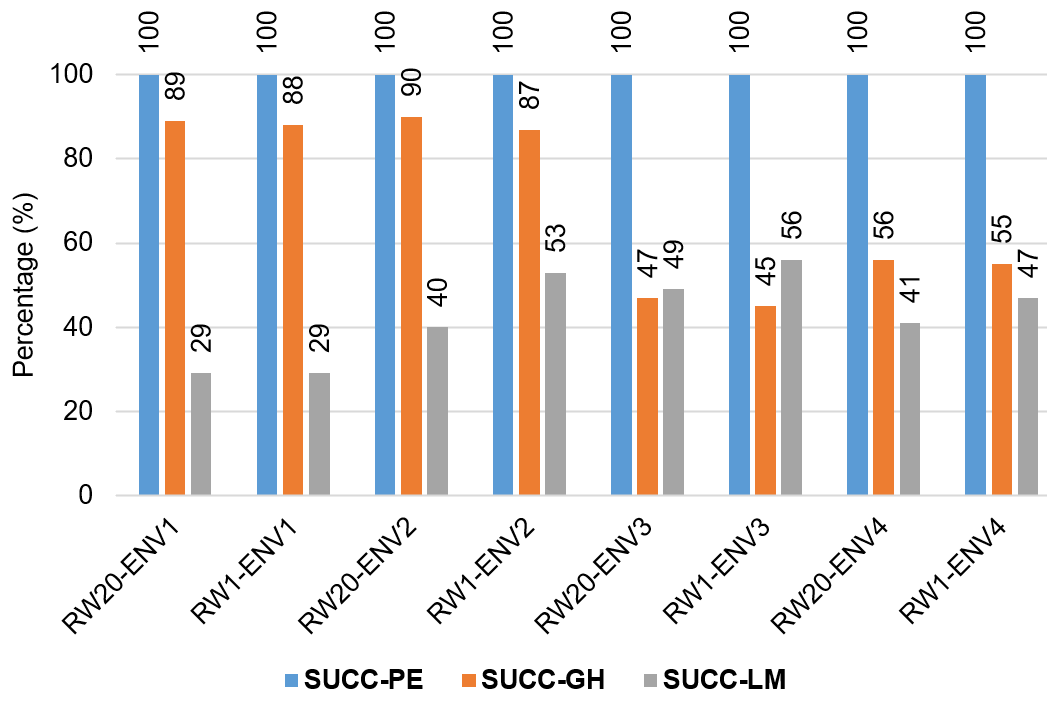}
	\label{}
}\\
\end{tabular}
\caption{The comparison of reward effectiveness.
}
\label{fig_reward-measure}
\end{figure}

\begin{figure}[!t] 
\centering
\begin{tabular}{cc}
\subfloat[\textbf{A2C Algorithm}]
{
	\includegraphics[width=0.46\textwidth]{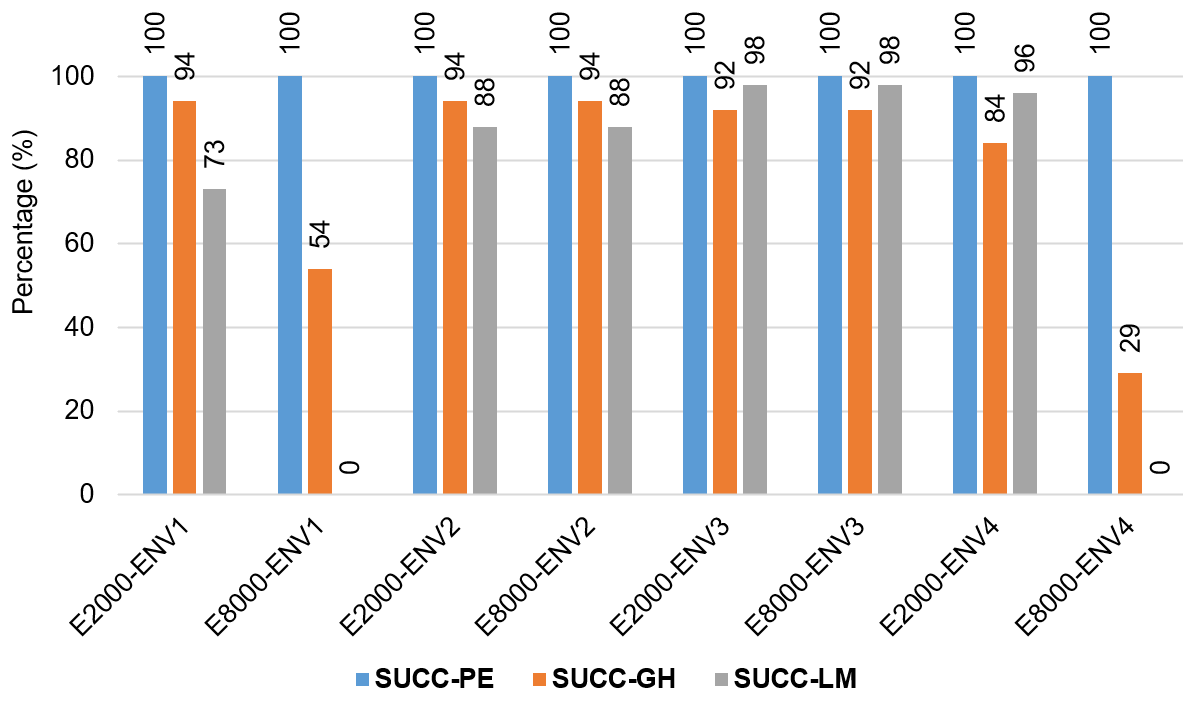}
	\label{}
}\\%
\subfloat[\textbf{PPO Algorithm}]
{
	\includegraphics[width=0.46\textwidth]{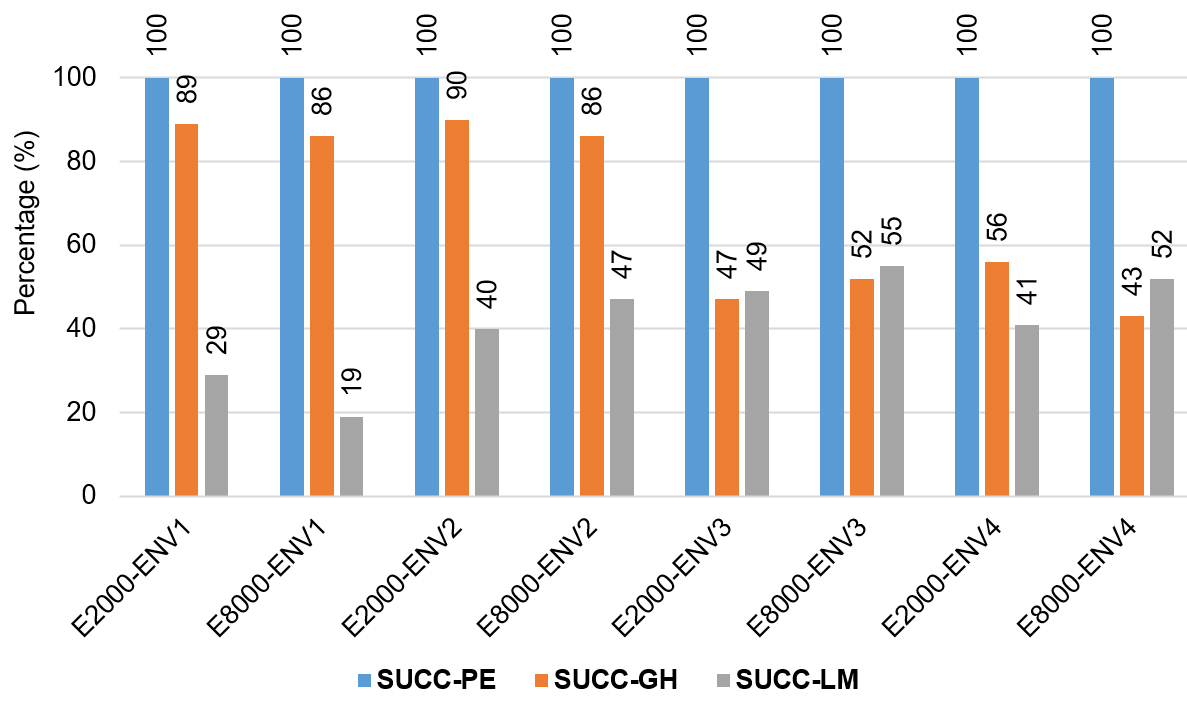}
	\label{}
}\\
\end{tabular}
\caption{The comparison of training episodes.}
\label{fig_effectiveness_episodes}
\end{figure}

\subsection{Experimental results}

\subsubsection{Time consumption of training process}

The time consumption for the learning process of A2C and PPO algorithms is shown in \textbf{Fig. \ref{time_consumption_training}}. The PPO algorithm takes much more time for training than A2C. With 2000 episodes, the training time of A2C consumes 40 minutes, while the consumption time of the PPO training process is higher by 15 minutes. For 4000 episodes and 8000 episodes, the difference in time consumed between the two algorithms is 40 minutes. Because the PPO has to compare the current policy with the previous one and save training parameters for the next comparison, it takes more time for the learning process than A2C. Moreover, the training process takes much more time if it is performed in a real environment, not to mention the amount of time for the environment reset.

\subsubsection{Evaluating effectiveness of rewards} \label{sec_eval_reward}
To clarify the effectiveness of rewards for RL model performance, we measure the values of rewards, including $r_s$ and $r_s$, in two cases:
\begin{itemize}
    \item RW1: $r_s$ = 1 for a correct action, $r_f$ = -1 for otherwise.
    \item RW20: $r_s$ = 20 for a correct action, $r_f$ = -1 for otherwise. In the training process, many actions are implemented to find the correct action. Therefore, the correct action should receive a higher reward value than the incorrect one. Besides, if there are correctly duplicate actions executed, only one receives the reward, and the others are unnecessary.
\end{itemize}

We train RL models with 2000 episodes using A2C and PPO algorithms to find the case of reward measurement that is optimal. Measurements are performed for the two cases of reward measurement (RW20 and RW1) in the four environments mentioned in \textbf{Section \ref{metrics_scenarios}} and the results are illustrated in \textbf{Fig. \ref{fig_reward-measure}}. Note that, each case of a specific pair of a reward measurement and an environment is indicated by \textit{X-Y}, where $X$ and $Y$ are reward and environment cases, respectively. For example, RW20-ENV1 is the result experienced with RW20 in ENV1.

As we can see from the charts, most of the results of RW20 in the four environments are better than those of RW1. \textit{SUCC-PE} of the two algorithms always hit the top (100\%). Considering \textit{SUCC-GH}, the percentage of RW20 is also always higher than that of RW1, especially in the result of A2C, which is a significant rise of the figure of RW20 compared to the formation of RW1. By contrast, the results of A2C measured by \textit{SUCC-LM} are considerably different between RW20 and RW1. Albeit less pronounced, with PPO, the figure of RW1 is better than that of RW20.

\subsubsection{Evaluating effectiveness of training episodes}
In this experiment, we train models of A2C and PPO algorithms with 2,000 episodes (E2000) and 8,000 episodes (E8000), then run 100 tests in the four environments (ENV1, ENV2, ENV3, and ENV4) with 4,000 steps at maximum as mentioned in \textbf{Section \ref{sec_evaluation_criteria}}. The SUCC-PE, SUCC-GH, and SUCC-LM metrics are used for evaluation. The reward RW20 is utilized for this experiment because of its optimization as considered in \textbf{Section \ref{sec_eval_reward}}.

\begin{figure*}[!t] 
\centering
\begin{tabular}{cc}
\subfloat[\textbf{ENV1}]
{
	\includegraphics[width=0.41\textwidth]{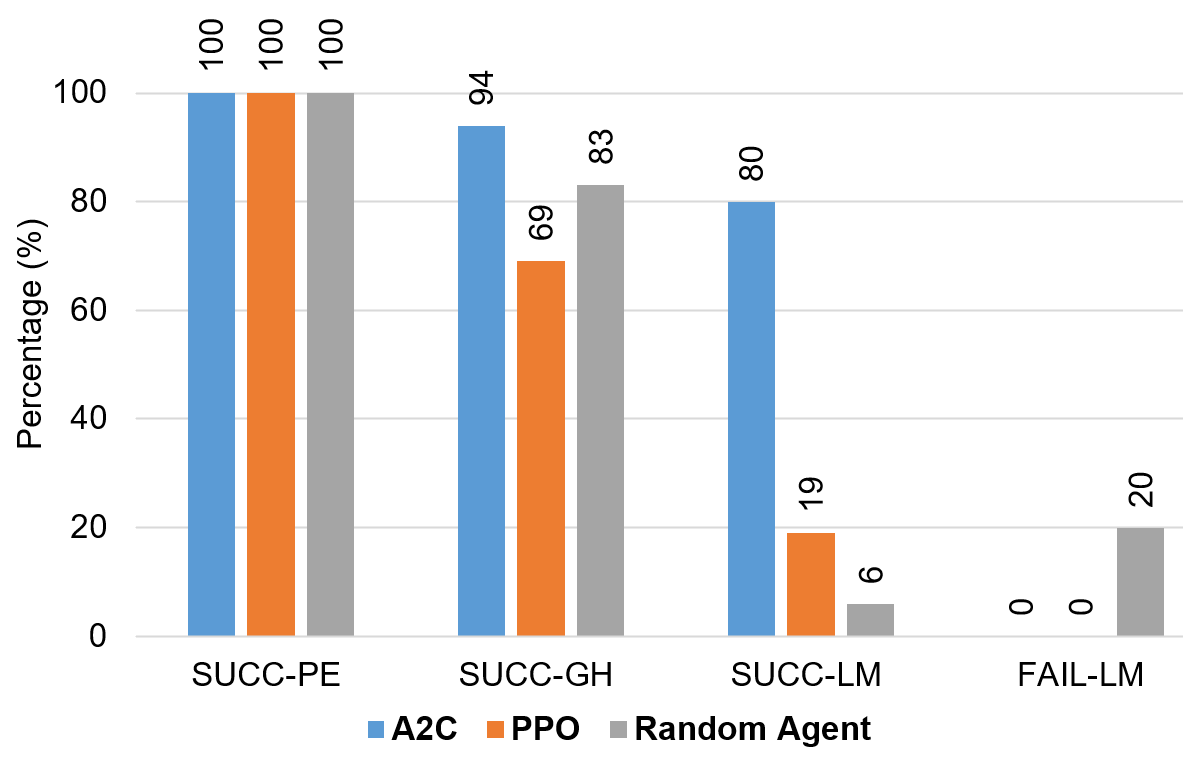}
	\label{}
}\hfill%
\subfloat[\textbf{ENV2}]
{
	\includegraphics[width=0.41\textwidth]{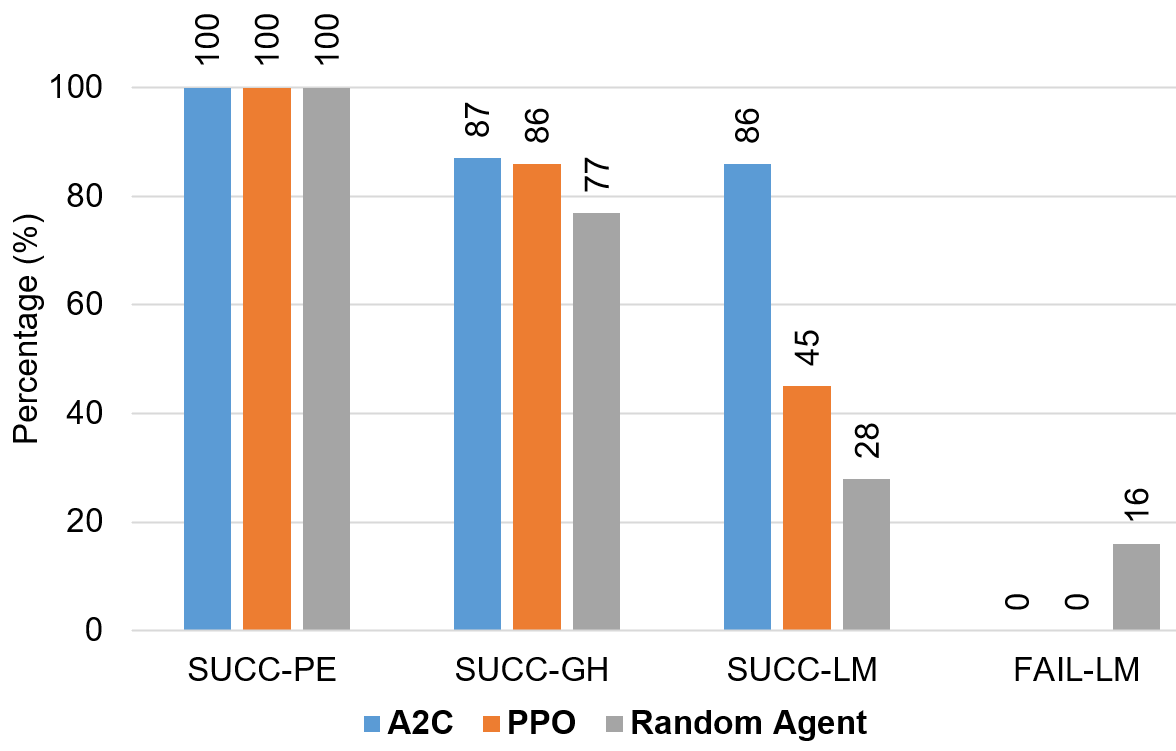}
	\label{}
}\\
\subfloat[\textbf{ENV3}]
{
	\includegraphics[width=0.41\textwidth]{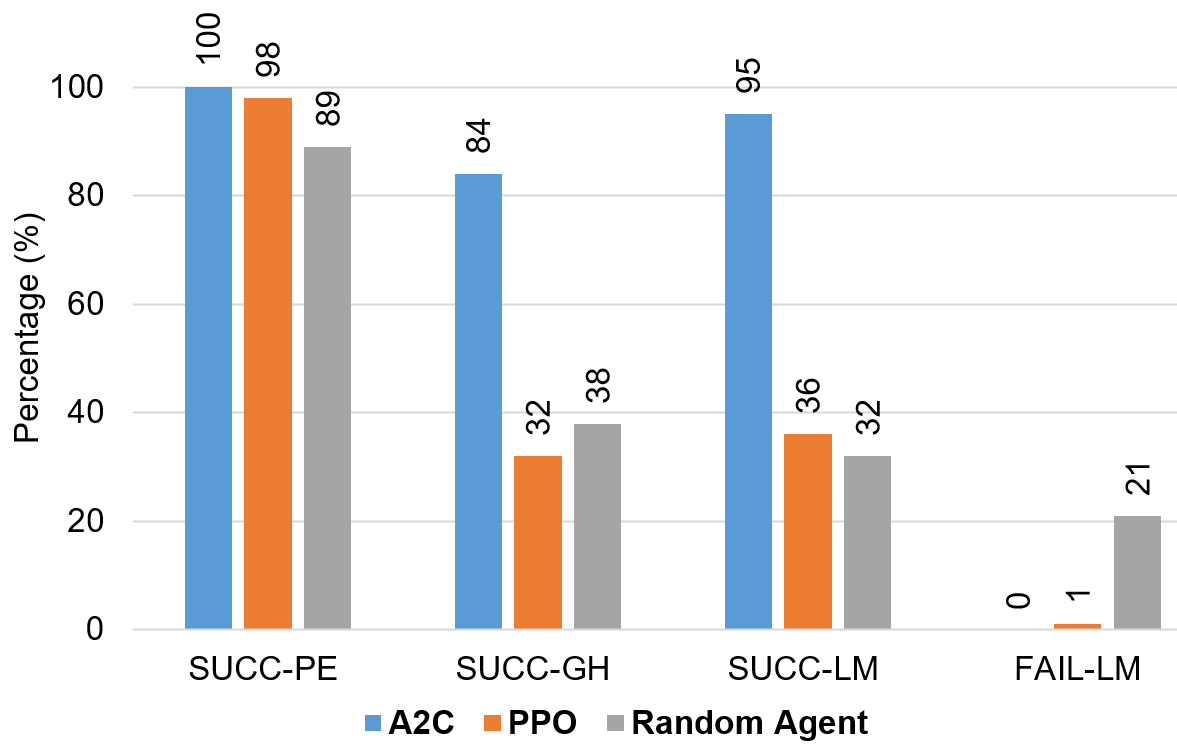}
	\label{}
}\hfill%
\subfloat[\textbf{ENV4}]
{
	\includegraphics[width=0.41\textwidth]{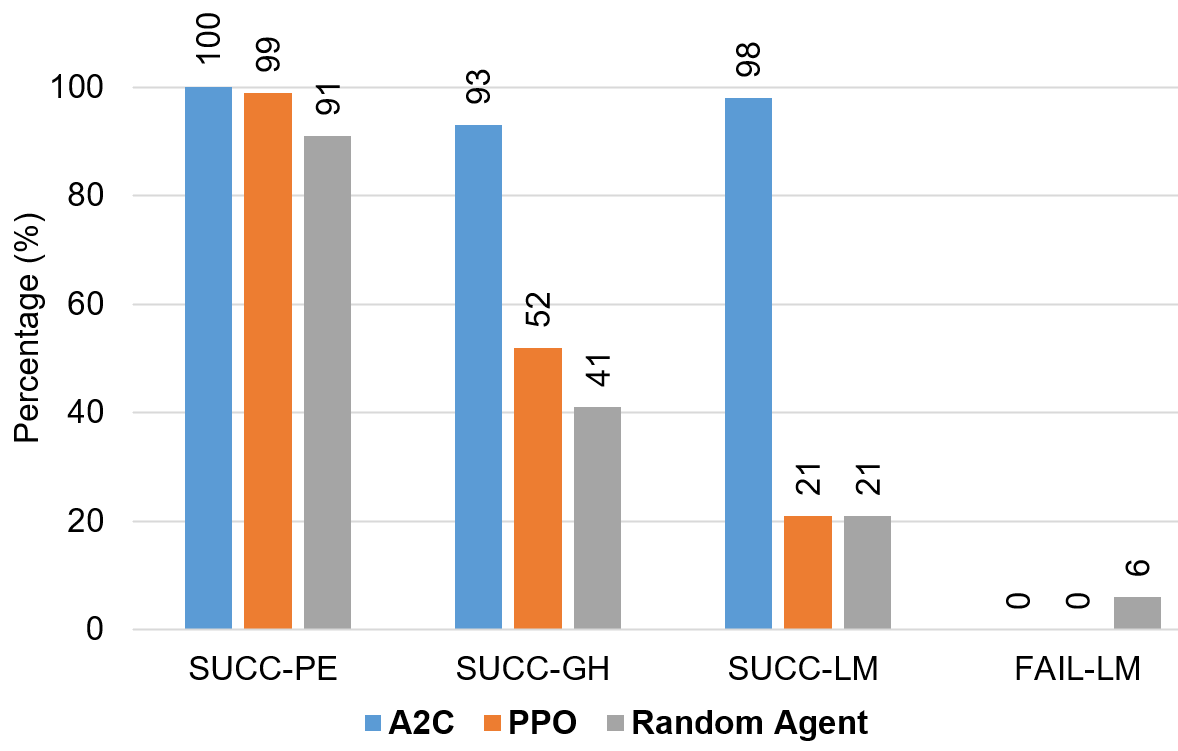}
	\label{}
}\\
\end{tabular}
\caption{The comparing successful attack ratio of the algorithms.
}
\label{comparison_effectiveness_algorithms}
\end{figure*}

\begin{figure*}[!t] 
\centering
\begin{tabular}{cc}
\subfloat[\textbf{ENV1}]
{
	\includegraphics[width=0.35\textwidth]{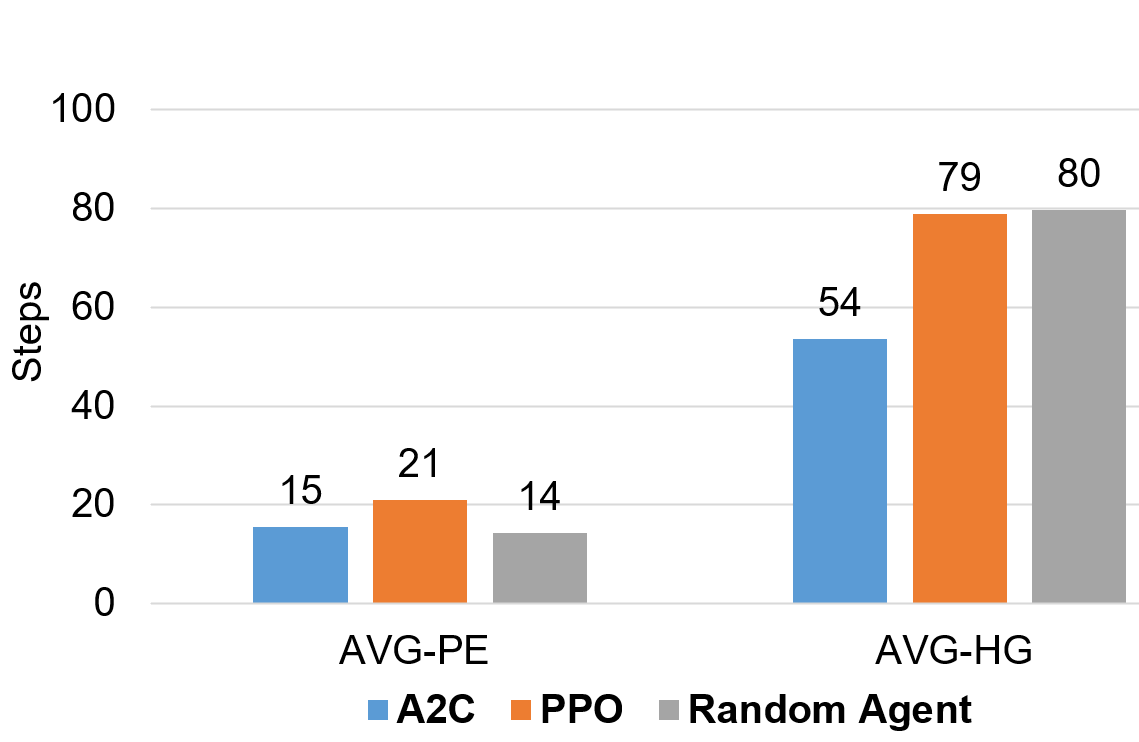}
	\label{}
}\hfill%
\subfloat[\textbf{ENV2}]
{
	\includegraphics[width=0.35\textwidth]{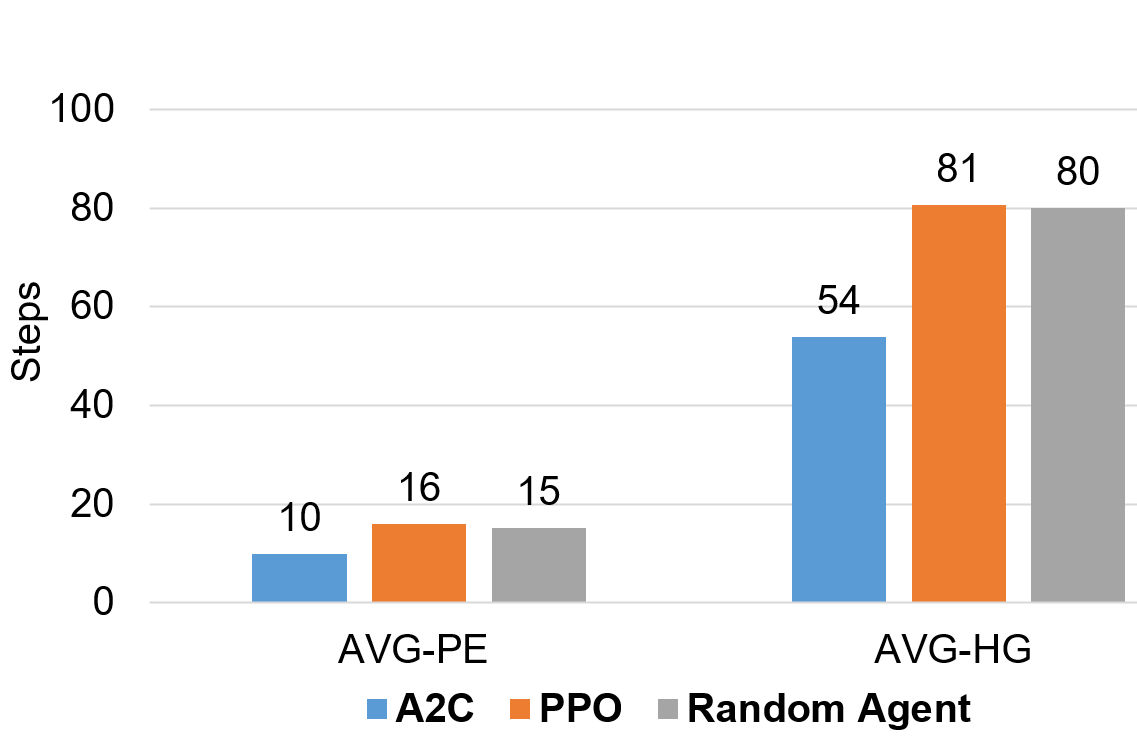}
	\label{}
}\\
\subfloat[\textbf{ENV3}]
{
	\includegraphics[width=0.35\textwidth]{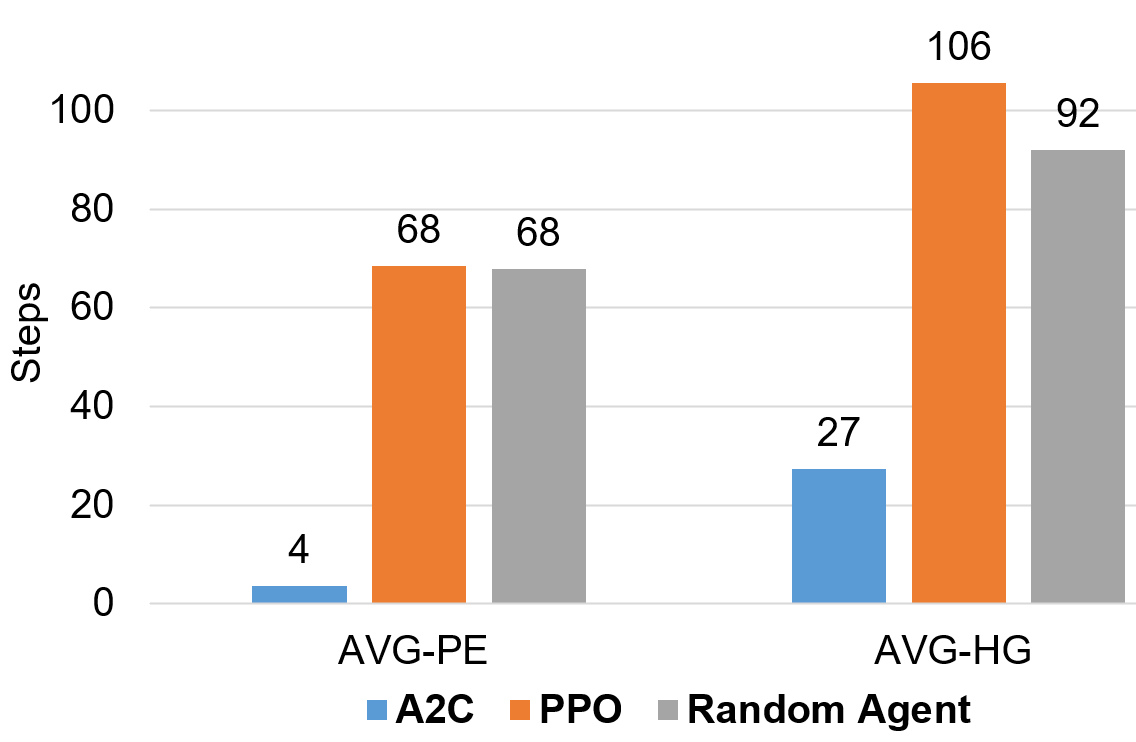}
	\label{}
}\hfill%
\subfloat[\textbf{ENV4}]
{
	\includegraphics[width=0.35\textwidth]{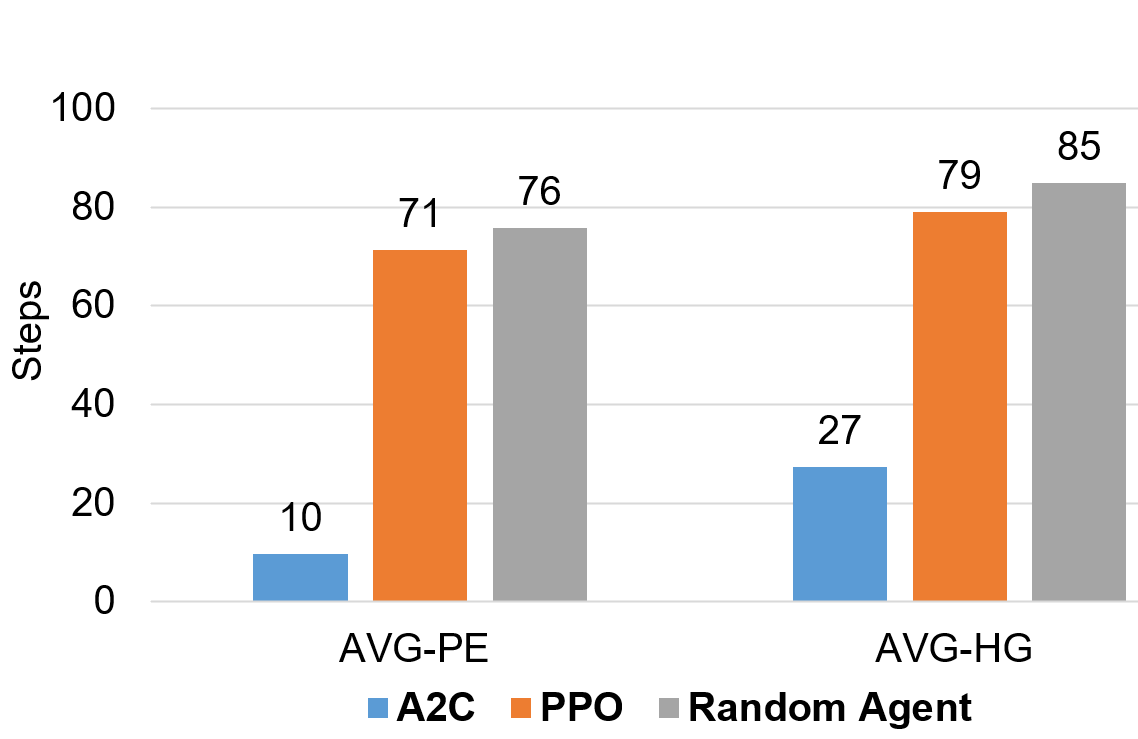}
	\label{}
}\\
\end{tabular}
\caption{The comparing the number of steps performed by various algorithms.
}
\label{comparison_effectiveness_algorithms_steps}
\end{figure*}

Results illustrated in \textbf{Fig. \ref{fig_effectiveness_episodes}} show that the models trained at 2,000 episodes give better values of metric than ones trained at 8000 episodes. Some results which are shown in \textbf{Fig. \ref{fig_effectiveness_episodes}-a} have a remarkable difference, for example, values of the SUCC-PE metric in ENV1 and ENV4. Considering the SUCC-LM of the A2C algorithm, the E2000-trained model gives a tremendous high at over 70 successful tests, whereas the figure given by the E8000-trained model is zero. However, the SUCC-LM metric measured by PPO has a contrast. The model trained with 8,000 episodes gives positive results compared to the E2000-trained one.

These results show that when increasing episodes for learning, the A2C algorithm loses effectiveness, which is shown in the metrics, especially in the values measuring LM attacks. By contrast, PPO gives relative stability (shown in \textbf{Fig. \ref{fig_effectiveness_episodes}-b}) given that there is little difference in results between the E2000 and E8000 trained models.

\subsubsection{Effectiveness comparison of A2C, PPO, and Random Agent algorithms}
To evaluate the effectiveness of various algorithms, this experiment conducts 100 tests with A2C, PPO, and Random Agent. The Random Agent (RA) is an algorithm that randomly selects actions in the action space. We limit the maximum of actions given in each test is 200 so that a test fails if it does not reach the target after 200 actions are selected. Moreover, agents of A2C and PPO are trained with 2000 episodes.

The \textbf{Fig. \ref{comparison_effectiveness_algorithms}} and \textbf{Fig. \ref{comparison_effectiveness_algorithms_steps}} depict the comparison of the three algorithms. Overall, the A2C algorithm produces a significant effect on both success levels and performed action numbers in the environments. Besides, the FAIL-LM metric of Random Agent is quite high in the environments, while that of A2C and PPO is almost non-existent.

With the SUCC-PE metric, the agents of three algorithms hit the top at the ratio of 100\% in ENV1 and ENV2, followed by a decrease in PPO and RA in ENV3 and ENV4. With SUCC-GH and SUCC-LM metrics, there is a significant difference between A2C and the others. The A2C algorithm always gives high performance compared to PPO and RA, which have roughly the same low performance. Besides, A2C and PPO are almost without fail in LM attacks, while RA gives a considerable fail ratio, especially in ENV1 and ENV3, hitting about 20\% in total.

Additionally, the average number of steps for launching PE and GH of A2C is more optimal than PPO and RA. It needs fewer steps for a successful attack, while PPO and RA have to perform more steps. For instance, the number of steps they take is approximately 5 times as many as A2C in ENV3 and ENV4.

\subsubsection{Comparing with related works}

\begin{table}[!b]
\caption{Compare with related works}
\label{table:compare_RW}
\resizebox{\linewidth}{!}{%
\begin{tabular}{ccccc}
\hline
\textbf{}                             & \textbf{}              & \textbf{Maeda et al. \cite{Ryusei_automating_post_exploitation_rl}} & \textbf{Kalle et al. \cite{Kujanp_automating_privilege_escalation_rl}} & \textbf{Ours} \\ \hline
\multirow{2}{*}{\textbf{Goal}}        & Privilege   escalation &                                                                                              & X                                                                                             & X             \\
                                      & Lateral movement       & X                                                                                            &                                                                                               & X             \\ \hline
\multirow{2}{*}{\textbf{Method}}      & A2C                    & X                                                                                            & X                                                                                             & X             \\
                                      & PPO                    &                                                                                              &                                                                                               & X             \\ \hline
\multirow{2}{*}{\textbf{Environment}} & PowerShell   Empire    & X                                                                                            & X                                                                                             &               \\
                                      & Metasploit             &                                                                                              &                                                                                               & X             \\ \hline
\multirow{2}{*}{\textbf{Target}}      & Windows                & X                                                                                            & X                                                                                             & X             \\
                                      & Linux                  &                                                                                              &                                                                                               & X             \\ \hline
\end{tabular}
}
\end{table}

In our work, we have conducted a comparative analysis, as shown in \textbf{Table \ref{table:compare_RW}}, between our approach and two related works by Maeda et al. \cite{Ryusei_automating_post_exploitation_rl} and Kalle et al. \cite{Kujanp_automating_privilege_escalation_rl}. While \cite{Ryusei_automating_post_exploitation_rl} and \cite{Kujanp_automating_privilege_escalation_rl} exclusively focus on privilege escalation or lateral movement, our analysis goes further by embracing a broader variety of strategies and targets. Notably, we perform experiments on various environments on both Linux and Windows platforms, providing a more comprehensive benchmarking results of our framework. Furthermore, we combine the system across different strategies, such as A2C and PPO, increasing the method's flexibility and efficiency. Significantly, our research opted not to create an interactive environment with PowerShell Empire, a framework that has long been deprecated and is no longer being developed or supported.

%% file: 2-Relatedwork.tex
\section{Related work} \label{sect_relatedworks}
\begin{table*}[!t]
\centering\
\caption{The summary of related works on PT}
\label{table:summary}
\resizebox{\textwidth}{!}{%
\def\arraystretch{1.3}%
\begin{tabular}{ccccc}
\hline
\textbf{Approach} &
  \textbf{Method} &
  \textbf{Environment type} &
  \textbf{Goal} &
  \textbf{Target} \\ \hline
Xue et al. \cite{autoPT_RW_qiu2014automatic} &
  Automatic graph generation &
  Static &
  Automated PT graph generation &
  Network   exploitable \\ \hline
Kyle et al. \cite{autoPT_RW_ingols2009modeling} &
  NetSPA &
  Static &
  \begin{tabular}[c]{@{}c@{}}Model network attacks \& \\ defenses using attack graphs\end{tabular} &
  Network   exploitable \\ \hline
Fabio et al. \cite{zennaro2023modelling} &
  tabular Q-Learning &
  Dynamic &
  Automating Capture the Flag (CTF) PT &
  CTF challenge \\ \hline
Isao \cite{deep_exploit} &
  A3C &
  Dynamic &
  \begin{tabular}[c]{@{}c@{}}Improving vulnerability diagnoses \\and exploitation tasks\end{tabular} &
  - \\ \hline
Yi and Liu \cite{Yi_DRL_pentesting} &
   mulVAL, Double DQN  &
  Dynamic &
  PT   attack path planning &
  Network   exploitable \\ \hline
Hu et al. \cite{hu2020automated} &
  mulVAL, DQN &
  Dynamic &
  Determining   the optimal attack path &
  Network   exploitable \\ \hline
Chaudhary et al. \cite{chaudhary2020automated} &
  DQN &
  Dynamic &
  \begin{tabular}[c]{@{}c@{}}Exploring a compromised network \\and finding sensitive documents\end{tabular} &
  Network   exploitable \\ \hline
Kalle et al. \cite{Kujanp_automating_privilege_escalation_rl} &
  A2C &
  Dynamic &
  Performing   privilege escalation &
  Windows \\ \hline
Maeda et al.   \cite{Ryusei_automating_post_exploitation_rl} &
  A2C &
  Dynamic &
  \begin{tabular}[c]{@{}c@{}}Performing   lateral movement \\and obtain domain controller privileges\end{tabular} &
  Windows \\ \hline
Mohamed et al. \cite{ghanem2023hierarchical} &
  Hierarchical RL &
  Dynamic &
  \begin{tabular}[c]{@{}c@{}}Automating decision-making\\ in PT tasks\end{tabular} &
  POMDP environment \\ \hline
Khuong et al. \cite{tran2022cascaded} &
  Cascaded RL &
  Dynamic &
  \begin{tabular}[c]{@{}c@{}}finding the optimal attack policy in PT scenarios \\with large discrete action spaces\end{tabular} &
  Network exploitable \\ \hline
\textbf{Raijū (Ours)} &
  \textbf{A2C, PPO} &
  \textbf{Dynamic} &
  \begin{tabular}[c]{@{}c@{}}\textbf{Performing privilege escalation, gathering}\\ \textbf{hashdump, and lateral movement}\end{tabular} &
  \textbf{Windows, Linux} \\ \hline
\end{tabular}%
}
\end{table*}

PT is a proactive method of identifying vulnerabilities in digital assets by actively seeking and exploiting weaknesses from the attacker's perspective.  By simulating invasive behaviors mirroring those of potential attackers, the process uncovers covert attack pathways within the system and subsequently assesses the system's security performance. There are several available PT tools such as Nmap \cite{orebaugh2011nmap}, Nessus \cite{rogers2011nessus}, Core Impact \cite{phong2014overview}, Caldera \cite{alford2022caldera}, and Metasploit \cite{metasploit}. Without scanning for changes to the environment occurring in real-time, Nessus and Nmap utilize the target list to test each network vulnerability individually. As a high-performing vulnerability exploit, Core Impact can automatically create an attack strategy on the target environment before performing PT. Similarly, Caldera automates adversarial action planning for Windows networks using an internal model of enterprise domains and attacker knowledge, guided by an intelligent heuristics-based planner. However, because those technologies rely on the expert judgment of security experts to make judgments, they frequently are unable to adapt attack plans as the environment changes. Even the powerful and well-known Metasploit Framework, which offers automation tools to find and exploit vulnerabilities in the target system, has restrictions on the modules and payload that can be used to exploit the vulnerability when PT is done manually.

From the aforementioned issues, researchers have developed automated PT techniques. Building automated pentest systems has become critical to ensure that pentest reports discoveries exhaustively test attack areas \cite{RW_cengiz2023reinforcement}.  Considering the reliability of attack paths, Xue et al. \cite{autoPT_RW_qiu2014automatic} introduced an algorithm that utilizes the Common Vulnerability Scoring System to automatically generate attack graphs for optimizing the network topology. The experimental results reveal its capability to generate multiple paths accurately and effectively. Utilizing the concept of attack graphs as a foundation, Kyle et al. \cite{autoPT_RW_ingols2009modeling}  introduced the NetSPA attack graph system, a novel approach that enables network defenders to evaluate security threats and choose complementary strategies. Notably, this system employs firewall rules and vulnerability scanning to swiftly analyze multiple targets, resulting in a substantial reduction in attack graph construction time. This work contributes significantly to the development of practical tools for network defense and security strategy optimization. Nevertheless, most of these approaches are capable only of generating predefined action sequences suitable for static environments. In simpler terms, they can offer retrospective instructions or guidelines for PT, yet lack the capability to conduct PT in a real-time and interactive manner within dynamic PT scenarios. Moreover, due to the requirement for comprehensive system understanding, the applicability of attack graphs to authentic dynamic penetration environments presents challenges. ML-based PT methods have emerged as a solution to the problems. 

Based on how the model learns, we can divide ML into 3 types: Supervised Learning, Unsupervised Learning, and RL. Among these, PT based on RL stands out as the most prevalent approach. The complexity and unpredictability of PT can be evaluated through the utilization of RL. RL effectively identifies optimal attack strategies but requires accurate exploitation modeling and realistic training simulators \cite{1_hoffmann2015simulated}. This is evident as modern ML heavily depends on extensive datasets, especially in supervised and semi-supervised learning \cite{survey_prudencio2023survey}. While creating large training datasets is crucial for automating security solutions, it's challenging in real-time, dynamic settings like post-exploitation. Consequently, supervised and unsupervised learning may not be the best choices for tasks in the PT field. In contrast, RL methods can learn directly from the environment, eliminating the need for pre-existing datasets \cite{adawadkar2022cyber}. In order for the taught RL agents to continuously do the best possible behaviors, they are modeled. The agents can thus be used in challenging real-time contexts.

For instance, the study of Fabio et al. \cite{zennaro2023modelling} focuses on automating complex PT tasks, specifically within the context of capture-the-flag (CTF) hacking challenges. It employs model-free RL algorithms to address these challenges and highlights the unique complexities associated with PT. To manage the exponential growth in complexity as the number of possible states and actions increases, the study emphasizes the use of prior knowledge to guide the learning process. By doing so, it aims to achieve more efficient and effective solutions for cybersecurity tasks, ultimately enhancing the automation of critical security assessments.

With the goal of conducting network security assessment, Hu et al. \cite{hu2020automated} introduced an automated PT framework that leverages Deep RL to streamline the PT process. In the initial phase, the framework can employ the Shodan search engine to gather relevant server information, thereby creating a realistic network topology. Subsequently, it utilizes multi-host multistage vulnerability analysis to generate an attack tree based on this topology. Conventional search techniques are then employed to identify all possible attack paths within the tree and construct a matrix representation required for Deep RL techniques. In the secondary phase, the framework employs the DQN technique to identify and exploit attack paths within feasible systems. In another study, Chaudhary et al. \cite{chaudhary2020automated} mentioned a different notion while applying the RL approach in the post-exploitation phase of PT. This study uses RL to train the agent by providing an appropriate environment for detecting sensitive documents and investigating compromised networks. In the study \cite{Yi_DRL_pentesting}, the MulVAL attack graph is integrated with the DDQN algorithm to enhance rewards using the prior knowledge from the attack graph. The experiments demonstrated that the MulVAL DDQN algorithm (MDDQN) accelerated the convergence speed and substantially improved the efficiency of attack path planning. The outcomes highlighted the increasingly evident benefits of the MDDQN algorithm as the experimental scenarios became more complex.

Addressing operating system exploitation, Kalle et al. \cite{Kujanp_automating_privilege_escalation_rl} introduced an RL-based approach for automated privilege escalation attacks within the Windows environment. Although their agent was trained in simulated settings, the test outcomes illustrate its ability to effectively solve the formalized privilege escalation problem on fully-featured Windows machines with authentic vulnerabilities, approaching optimal performance. In another study, Ryusei Maeda et al. \cite{Ryusei_automating_post_exploitation_rl} also explored automatic exploitation using deep RL, focusing on the post-exploit phase. Their deep RL agent advances adversarial emulation in real environments by specializing in Windows domain lateral movement. They trained the agent with PowerShell Empire modules as the action space, and its state includes key factors like network-discovered computers, compromised systems, and local administrative privileges. The authors show the agent's ability to learn and execute lateral movement, ultimately gaining domain controller privileges. Moreover, at the Black Hat USA 2018 Arsenal event, Isao Takaesu presented DeepExploit \cite{deep_exploit}, an RL-powered automated exploitation tool designed to exploit server vulnerabilities. DeepExploit, based on the Metasploit framework, is a deep RL agent specialized in automating initial access through known vulnerabilities and exploits. Following successful penetration, it attempts recursive access to other hosts within the specified input IP address's local network.

Taking a contextual perspective, where PT tasks are treated as partially observed Markov decision processes (POMDPs), Mohamed et al. \cite{ghanem2023hierarchical} introduce the Intelligent Automated Penetration Testing Framework (IAPTF). This framework employs Hierarchical RL to automate decision-making in PT tasks and addresses the challenge of solving large POMDPs in extensive networks through a hierarchical network modeling approach. Notably, this advantage scales with the size of the network. Furthermore, IAPTF offers the added benefit of simplifying the repetition of testing similar networks, a common occurrence in real-world PT scenarios. 
In another study, Khuong et al. \cite{tran2022cascaded} described a novel architecture called deep Cascaded Reinforcement Learning Agents (CRLA), which was created to address the issues faced by huge discrete action spaces in autonomous PT simulations. CRLA outperforms conventional DQN agents often used in autonomous PT by determining optimum attack tactics in complex cybersecurity networks via an algebraic action decomposition. The study confirms that CRLA performs better than single DDQN agents in simulated CybORG settings, demonstrating its competitive advantage. Additionally, CRLA's potential for use in real-world scenarios is highlighted by its scalability to larger action areas with sub-linear computing complexity.

Various related works have significantly contributed to the advancement of PT, with each project offering unique insights and innovations. However, several of these studies face notable limitations. For instance, certain studies, such as \cite{Ryusei_automating_post_exploitation_rl, deep_exploit}, encounter challenges when utilizing outdated tools like PowerShell \cite{Ryusei_automating_post_exploitation_rl} or require further modifications to function optimally \cite{deep_exploit}. Addressing these issues demands a considerable amount of technical expertise and effort, both for initial development and future expansion. Moreover, some works, like \cite{Kujanp_automating_privilege_escalation_rl, Ryusei_automating_post_exploitation_rl}, narrow their focus to the Windows platform environment, overlooking the prevalence of other commonly used platforms like Linux in contemporary network systems. Additionally, the aforementioned related studies predominantly tackle individual aspects in isolation, lacking the interconnections that could enhance flexibility. This background acts as a motivation for our study and project development. A comprehensive summary of the aforementioned related studies can be found in \textbf{Table~\ref{table:summary}}.

%% file: 6-Conclusion.tex
\section{Conclusion} \label{sec_conclusion}
In this paper, we have addressed the critical task of assessing the risks inherent in information systems, specifically focusing on the post-exploitation phase after successful attacks. While there exist several efficient tools to aid in post-exploitation activities, the automation of this intricate process remains a challenge. Often, the responsibility falls upon seasoned security experts, commonly known as penetration testers or pen-testers, to execute these steps with their extensive security knowledge. Our research introduces Raijū, an innovative automation approach driven by RL, tailored to expedite the post-exploitation process. Raijū acts as a supportive tool, assisting pen-testers in swiftly implementing post-exploitation procedures for robust security-level evaluations within network systems. We extend our methodology to leverage the potent exploit modules of the Metasploit framework, employed to launch attacks during penetration tests.

To empirically evaluate the efficacy of our proposed approach, we employ two RL algorithms, A2C and PPO, to train agents. These agents assume the role of navigating server states, meticulously seeking potential avenues for exploiting vulnerabilities. These specialized agents, designed to automate privilege escalation, hashdump gathering, and lateral movement using Metasploit modules, have achieved a remarkable success rate of over 84\% in real environments, accomplishing their tasks in fewer than 55 attack steps. Notably, A2C has consistently outperformed PPO and RA, demonstrating its effectiveness in selecting optimal actions for post-exploitation automation. With fewer steps required for successful attacks, A2C offers a more efficient and responsive approach to penetration testing automation, enhancing our ability to address emerging threats and vulnerabilities.

The outcomes of our study bear significant implications for the realm of security assessment. Raijū's integration of RL empowers pen-testers with an advanced tool capable of augmenting their expertise in swiftly executing the post-exploitation process. The success demonstrated by our proposed approach underscores the potential of automation in enhancing the precision and efficiency of security evaluations within network systems. As the landscape of cybersecurity continues to evolve, approaches such as Raijū pave the way for innovative solutions that not only streamline processes but also empower security professionals in their pursuit of safeguarding information systems.

\section*{Acknowledgement}
This research was supported by The VNUHCM-University of Information Technology's Scientific Research Support Fund.